\documentclass[twocolumn,aps,showpacs,showkeys,tightenlines,superscriptaddress]{revtex4}
\usepackage{epsfig,bm}
\newcommand{\vi}[1]{\mbox{\boldmath $#1$}}
\newcommand{\vis}[1]{\mbox{\boldmath ${\scriptstyle #1}$}}
\begin{document}

\title{Proton-nucleus total reaction cross sections in the optical limit
Glauber theory: Subtle dependence on the equation of state of nuclear matter}

\author{K. Iida}
\affiliation{Department of Natural Science, 
Kochi University, Akebono-cho, Kochi 780-8520, Japan}
\affiliation{RIKEN Nishina Center, RIKEN, Wako-shi, 
Saitama 351-0198, Japan}
\author{K. Oyamatsu}
\affiliation{RIKEN Nishina Center, RIKEN, Wako-shi, 
Saitama 351-0198, Japan}
\affiliation{Department of Human Informatics, 
Aichi Shukutoku University, Nagakute, Nagakute-cho, 
Aichi-gun, Aichi 480-1197, Japan}
\author{B. Abu-Ibrahim}
\affiliation{RIKEN Nishina Center, RIKEN, Wako-shi, 
Saitama 351-0198, Japan}
\affiliation{Department of Physics, Cairo University, 
Giza 12613, Egypt}
\author{A. Kohama}
\affiliation{RIKEN Nishina Center, RIKEN, Wako-shi, 
Saitama 351-0198, Japan}

\begin{abstract}
     We calculate the proton-nucleus total reaction cross sections  
at different energies of incident protons
within the optical limit approximation of the 
Glauber theory.  The isospin effect has been taken into account. 
The nucleon distribution is obtained in the framework of macroscopic 
nuclear models in a way depending on the equation of state of uniform 
nuclear matter near the saturation density.  We find that
at an energy of order 40 MeV, the reaction cross section calculated 
for neutron-rich isotopes significantly increases as the parameter $L$ 
characterizing the density dependence of the symmetry energy increases,
while at energies of order 300 and 800 MeV, it is almost independent
of $L$.  This is a feature of the optical limit Glauber theory in which
an exponential dependence of the reaction cross section on the neutron
skin thickness remains when the total proton-neutron cross section
is small enough.

\end{abstract}

\maketitle

\section{Introduction}
     Reactions of unstable neutron-rich nuclei with a proton target
are of current interest as such reactions can act as a
major means to probe the matter densities of exotic nuclei.
If one appropriately selects incident energies, protons could be more 
sensitive to neutron distributions than to proton distributions of nuclei.
The equation of state (EOS) of nuclear matter is also of interest 
since it is essential to understand the saturation property of atomic 
nuclei, the structure of neutron stars, and the mechanism of 
stellar collapse.  In a previous work \cite{IOA} we have theoretically 
connected the EOS of nuclear matter to the proton-nucleus elastic 
differential cross sections.  We limited ourselves to high energy of 
order 800 MeV and used the optical limit approximation (OLA) of the 
Glauber theory.  We find, at large neutron excess, a strong 
correlation between the peak angle in the small momentum 
transfer regime and the density symmetry coefficient $L$, which 
characterizes the density dependence of the symmetry energy and hence 
controls the size of neutron-rich nuclei \cite{OI}.  At this high energy, we 
briefly addressed the question of how the total reaction cross sections and 
the EOS parameters connect and did not find a strong correlation 
as compared with the case of elastic scattering.

     The purpose of the present paper is to study the connection between 
the EOS of nuclear matter and the proton-nucleus total reaction cross 
section in more detail and at different energies within the OLA of the
Glauber theory.  We are interested 
in the total reaction cross section because, for neutron-rich isotopes,
it is much easier to measure experimentally than the elastic differential 
cross section and uncertainties in the measured reaction cross section
are expected to be sufficiently small to deduce new information about
the EOS parameters.

     In the course of this study, however, we happened to find some
interesting properties of the calculated total reaction cross sections, 
of which the physical manifestation remains to be examined.  We will thus 
focus on these properties from a purely theoretical point of view, instead 
of deducing the EOS parameters from empirical data for the total
reaction cross sections.

     In Sec.\ II, we summarize a macroscopic nuclear model that allows
us to connect the nucleon density distributions and the EOS of nuclear matter.
Section III is devoted to description of the reaction cross sections within 
the framework of the Glauber theory.  In Sec.\ IV, we calculate
the reaction cross sections for heavy neutron-rich nuclei in a way dependent 
on the EOS of nuclear matter, which in turn are analyzed by using the
rectangular distributions that allow for neutron skin thickness.

\section{From Equation of State to Nuclei}

    In this section, we summarize the way we connect the nucleon density
distributions with the EOS of nuclear matter.
Generally, the energy per nucleon of nuclear matter can be expanded around the
saturation point of symmetric nuclear matter as
\begin{equation}
    w=w_0+\frac{K_0}{18n_0^2}(n-n_0)^2+ \left[S_0+\frac{L}{3n_0}(n-n_0)
      \right]\alpha^2.
\label{eos0}
\end{equation}
Here $w_0$, $n_0$, and $K_0$ are the saturation energy, the saturation density 
and the incompressibility of symmetric nuclear matter, $n$ is the nucleon
density, and $\alpha$ is the neutron excess.  The parameters $L$ and 
$S_0$ characterize the density dependent symmetry energy coefficient $S(n)$: 
$S_0$ is the symmetry energy coefficient at $n=n_0$, and 
$L=3n_0(dS/dn)_{n=n_0}$ is the density symmetry coefficient.  As shown 
in Ref.\ \cite{OI} by describing macroscopic
nuclear properties in a manner that is dependent on these EOS parameters, 
empirical data for masses and radii of stable nuclei can provide a strong 
constraint on the parameters $w_0$, $n_0$, and $S_0$, while leaving $K_0$ and 
$L$ uncertain.  We remark that the isoscalar giant monopole resonance in 
nuclei (e.g., Ref.\ \cite{YCL}) and caloric curves in nuclear collisions 
(e.g., Ref.\ \cite{Nat}) can constrain $K_0$ only in a way that is dependent 
on models for the effective nucleon-nucleon interaction.

     The incompressibility $K_0$ and the density symmetry coefficient $L$ 
control in which direction the saturation point moves on the density versus 
energy plane, as the neutron excess increases from zero.  This feature can be 
found from the fact that up to second order in $\alpha$, the saturation energy
$w_s$ and density $n_s$ are given by 
\begin{equation}
  w_s=w_0+S_0 \alpha^2
\label{ws}
\end{equation}
and
\begin{equation}
  n_s=n_0-\frac{3 n_0 L}{K_0}\alpha^2.
\label{ns}
\end{equation}
The influence of $K_0$ and $L$ on neutron star structure can be significant
\cite{Prakash} despite the fact that these parameters characterize
the EOS near normal nuclear density and proton fraction, which are fairly 
low and large, respectively, as compared with the typical densities and
proton fractions in the central region of the star.

      We proceed to macroscopic nuclear models used in this and previous
works \cite{OI,IOA,OI2}.  We describe a spherical nucleus of proton number 
$Z$ and mass number $A$ within the framework of a simplified version of the 
extended Thomas-Fermi theory \cite{O}.  We first write the total energy of 
a nucleus as a function of the neutron and proton density distributions 
$n_n({\vi r})$ and $n_p({\vi r})$ in the form
\begin{equation}
 E=E_b+E_g+E_C+(A-Z)m_n c^2+Zm_p c^2,
\label{e}
\end{equation}
where 
\begin{equation}
  E_b=\int d {\vi r} n({\vi r})w\left(n_n({\vi r}),n_p({\vi r})\right)
\label{eb}
\end{equation}
is the bulk energy with the energy per nucleon $w(n_n,n_p)$ of uniform nuclear
matter,
\begin{equation}
  E_g=F_0 \int d {\vi r} |\nabla n({\vi r})|^2
\label{eg}
\end{equation}
is the gradient energy with adjustable constant $F_0$,
\begin{equation}
  E_C=\frac{e^2}{2}\int d {\vi r} \int  d {\vi r'} 
      \frac{n_p({\vi r})n_p({\vi r'})}{|{\vi r}-{\vi r'}|}
\label{ec}
\end{equation}
is the Coulomb energy, and $m_n$ and $m_p$ are the neutron and proton rest
masses.  We express $w$ as \cite{O}
\begin{equation}
  w=\frac{3 \hbar^2 (3\pi^2)^{2/3}}{10m_n n}(n_n^{5/3}+n_p^{5/3})
      +(1-\alpha^2)\frac{v_s(n)}{n}+\alpha^2 \frac{v_n(n)}{n},
\label{eos1}
\end{equation}
where 
\begin{equation}
  v_s=a_1 n^2 +\frac{a_2 n^3}{1+a_3 n}
\label{vs}
\end{equation}
and
\begin{equation}
  v_n=b_1 n^2 +\frac{b_2 n^3}{1+b_3 n}
\label{vn}
\end{equation}
are the potential energy densities for symmetric nuclear matter and pure 
neutron matter.  This expression for $w$ is one of the simplest 
parametrization that reduces to Eq.\ (\ref{eos0}) in the simultaneous limit of
$n\to n_0$ and $\alpha\to0$.  We then set the nucleon distributions $n_i(r)$ 
$(i=n,p)$ as
\begin{equation}
  n_i(r)=\left\{ \begin{array}{lll}
  n_i^{\rm in}\left[1-\left(\displaystyle{\frac{r}{R_i}}\right)^{t_i}\right]^3,
         & \mbox{$r<R_i,$} \\
             \\
         0,
         & \mbox{$r\geq R_i,$}
 \end{array} \right.
\label{ni}
\end{equation}
and in the spirit of the Thomas-Fermi approximation minimize the total 
energy $E$ with respect to $R_i$, $t_i$, and $n_i^{\rm in}$ with the mass 
number $A$, the EOS parameters $(n_0,w_0,S_0,K_0,L)$ and the gradient 
coefficient $F_0$ fixed.  By calculating the charge number, mass excess, and 
root-mean-square charge radius from the minimizing values of $R_i$, $t_i$, and 
$n_i^{\rm in}$ and fitting the results to the empirical values for stable 
nuclei $(25\leq A \leq 245)$ on the smoothed beta stability line, we finally 
obtain $n_0$, $w_0$, $S_0$, and $F_0$ for various sets of $L$ and $K_0$ 
ranging 0 MeV $<L<175$ MeV and 180 MeV $\leq K_0 \leq 360$ MeV.  We 
remark that as a result of this fitting, the parameters $a_1$--$b_2$ become 
functions of $K_0$ and $L$, while we fix the remaining parameter $b_3$, which
controls the EOS of matter for large neutron excess and high density, at 
1.58632 fm$^3$ throughout this fitting process.

     The macroscopic nuclear models used here can describe gross nuclear 
properties such as masses and root-mean-square radii in a manner that is 
dependent on the EOS parameters, $L$ and $K_0$.  Notably, as in Ref.\ 
\cite{OI}, these models predict that the root-mean-square matter radii increase
appreciably with $L$, while being almost independent of $K_0$, and that
the root-mean-square charge radii are almost independent of both $L$ and 
$K_0$.  However, there 
are some limitations in the present macroscopic approach.  First, this 
approach works well in the range of $\alpha\lesssim0.3$ and $A\gtrsim50$, 
where a macroscopic view of the system is relevant.  Second, the nuclear 
surface is not satisfactory in the present Thomas-Fermi-type theory, which 
tends to underestimate the surface diffuseness and does not allow for the 
tails of the nucleon distributions.  
Lastly, no shell and pairing effects are included.

\section{Glauber model}

     In the OLA of the Glauber model~\cite{Glauber}, the proton-nucleus 
scattering phase-shift 
function is simply described by the proton density $n_{p}({\vi r})$
and the neutron density $n_{n}({\vi r})$ as follows:
\begin{equation}
   {\rm e}^{i\chi_{\rm OLA}({\vis b})}
   = \exp \left[i\chi_{p}({\vi b})+
                i\chi_{n}({\vi b}) \right],
\label{ola}
\end{equation}
where $\chi_{p}$ ($\chi_{n}$) is the phase shift due to protons 
(neutrons) inside the nucleus, 
\begin{eqnarray}
i\chi_{p}({\vi b}) &=& -{\int{d{\vi r} }} n_{p}({\vi r})
                        \Gamma_{pp}({\vi b} - {\vi s}), 
\nonumber\\
i\chi_{n}({\vi b}) &=& -{\int{d{\vi r} }} n_{n}({\vi r})
                        \Gamma_{pn}({\vi b}-{\vi s}),
\label{chi}
\end{eqnarray}
with the impact parameter ${\vi b}$ and
the projection ${\vi s}$ of the coordinate ${\vi r}$ on a
plane perpendicular to the incident proton momentum.

     The profile function, $\Gamma_{pN}$, for $pp$ and $pn$ scatterings, 
is usually parametrized in the form:
\begin{equation}
  \Gamma_{pN}({\vi b}) =
  \frac{1-i\alpha_{pN}}{4\pi\beta_{pN}}\,\,
  \sigma_{pN}^{\rm tot}\, {\rm e}^{-{\vis b}^2 /(2\beta_{pN}) },
\label{gfn}
\end{equation}
where $\alpha_{pN}$ is the ratio of the real to the imaginary part of 
the $pp$ ($pn$) scattering amplitude in the forward direction, 
$\sigma_{pN}^{\rm tot}$ is the $pp$ ($pn$) total cross section, 
and $\beta_{pN}$ is the slope parameter of the $pp$ ($pn$) elastic 
scattering differential cross section.  These parameters are tabulated 
for different incident energies in Ref.~\cite{bhks}.

     The total reaction cross section of the proton-nucleus collision is 
calculated from 
\begin{equation}
  \sigma_{\rm R} = \int{d{\vi b}\,
  \left(1-\big|{\rm e}^{i\chi_{\rm OLA}({\vis b})}\big|^{2}\right)}.
\label{reaccs}
\end{equation}

\section{EOS dependence of reaction cross sections}

     In our calculation we have used a code based on the Monte Carlo 
integration for evaluations of the phase shift function, which can be 
applied to any arbitrary form of nucleon distributions.  We have 
calculated the total reaction cross section for $p$-$^{63}$Cu, 
$p$-$^{80}$Ni, $p$-$^{112}$Sn, $p$-$^{124}$Sn, and $p$-$^{208}$Pb by 
substituting the distributions (\ref{ni}) determined as functions of
$L$ and $K_0$ into $n_{n,p}({\vi r})$ in Eq.\ (\ref{chi}).
For each system at each energy we have performed summation over 228 
random points, which ensures sufficient accuracy.

\begin{figure}[t]
\epsfig{file=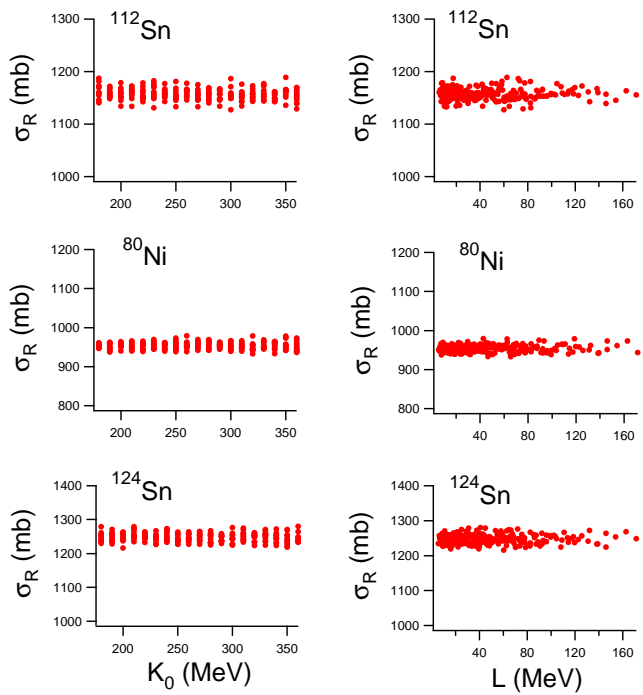,scale=0.54}
\caption{(Color online) The total reaction cross sections calculated
as a function of $K_{0}$ and $L$ for $p$-$^{112,124}$Sn 
and $p$-$^{80}$Ni at 800 MeV.}
\label{800.reac}
\end{figure}

\begin{figure}[b]
\epsfig{file=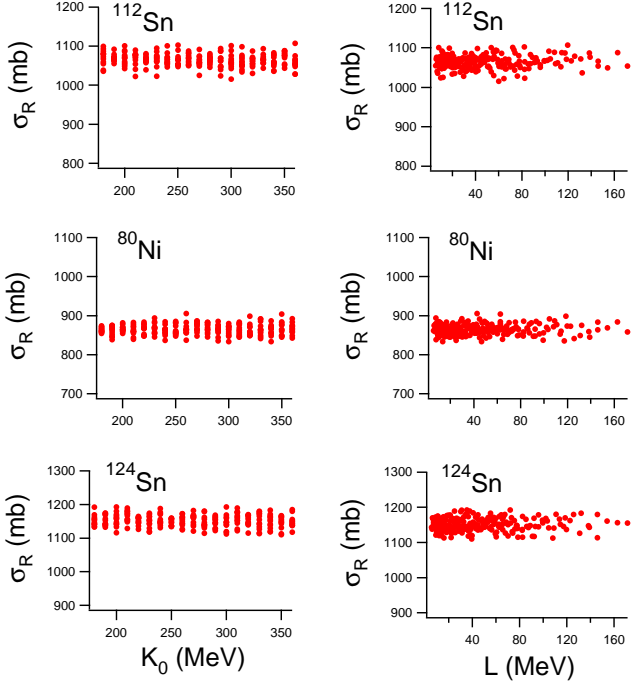,scale=0.54}
\caption{(Color online) The total reaction cross sections calculated
as a function of $K_{0}$ and $L$ for $p$-$^{112,124}$Sn 
and $p$-$^{80}$Ni at 300 MeV.}
\label{300.reac}
\end{figure}

     Figure~\ref{800.reac} shows the results for the selected isotopes 
at 800 MeV.  As we pointed out in our previous paper \cite{IOA} 
the dependence of 
the total reaction cross section on the EOS parameters is rather weak.
By decreasing the energy to 300 MeV, we obtain the same conclusion 
as at 800 MeV; the results are shown in Fig.~\ref{300.reac}.
Note that at such high energies the reaction cross section does not 
obey an empirical law obtained within a black sphere approximation 
\cite{BS2}, which suggests that the reaction cross section increases with 
the nuclear size and thus $L$.  This is a feature to be discussed in more 
details below.

\begin{figure}
\epsfig{file=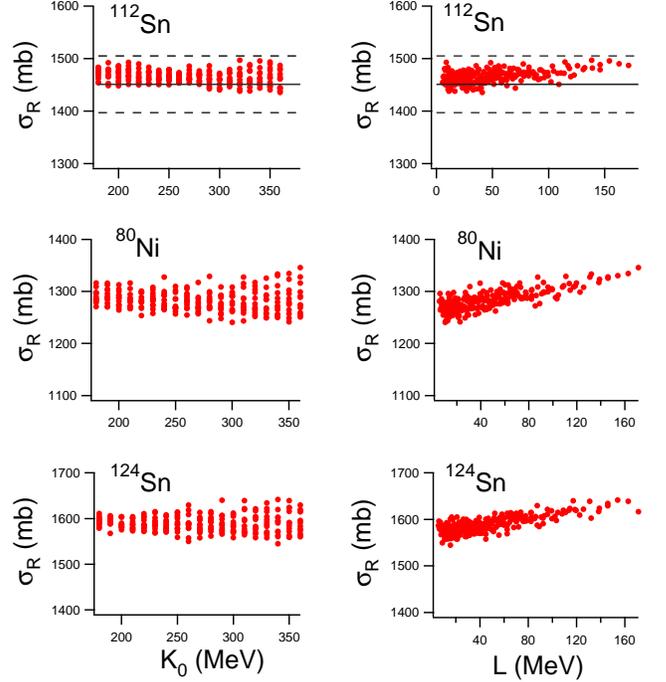,scale=0.54}
\caption{(Color online) The total reaction cross sections calculated
as a function of $K_{0}$ and $L$ for $p$-$^{112,124}$Sn 
and $p$-$^{80}$Ni at 40 MeV. 
The experimental values (solid lines: central values, dashed lines: 
upper and lower bounds) are taken from Ref.~\cite{carlson}.}
\label{40.reac}
\end{figure}

     At 40 MeV the $\sigma^{\rm tot}_{pn}$ and $\sigma^{\rm tot}_{pp}$ are 
fairly large compared to those at 800 MeV, and $\sigma^{\rm tot}_{pn}$ is
much larger than $\sigma^{\rm tot}_{pp}$.   Therefore we expect for neutron 
rich nuclei that the total reaction cross section depends sensitively on the 
neutron density distribution including the skin region.

     Figure~\ref{40.reac} shows our calculations of the reaction cross 
section as a function of $L$ and $K_{0}$.  We find, for neutron rich isotopes 
like $^{80}$Ni ($Z/A=0.35$), a strong dependence on the $L$ value.
The value of the total reaction cross section for ($L$,$K_{0}$)=(12.4, 350) 
in MeV is 1248 mb, while that for ($L$,$K_{0}$)=(171.3, 360) in MeV is 
1346 mb.  The difference between them is about 7$\%$.  If uncertainties 
of experimental data for the reaction cross section are less than about 
3.5$\%$ and our calculations with simplified scattering theory and density 
distributions are satisfactory, we could determine the $L$ value from
comparison between the calculated and measured values.

     For possible determination of $L$, therefore, it is important to 
examine how 
well the OLA of the Glauber model predicts the reaction cross section.  
Although this depends on the choice of density distributions,  for 
simplicity, we here confine ourselves to rectangular nucleon distributions, 
for which one can calculate the reaction cross section analytically 
in the zero range limit ($\beta_{pN}\to0$).

     Let us set the radius and density of the proton (neutron) distribution
for a nucleus of given $A$ and $Z$ to be $a_p$ ($a_n$) and
$n_{p0}=3Z/4\pi a_p^3$ [$n_{n0}=3(A-Z)/4\pi a_n^3$].  If $a_p=a_n\equiv a$, 
one can calculate the reaction cross section (\ref{reaccs}) in the zero range
limit as
\begin{equation}
  \sigma_{\rm R} = \pi a^2 \left[1-\frac{2}{\xi^2}
   +\frac{2}{\xi}\left(1+\frac{1}{\xi}\right)e^{-\xi}\right],
\label{reacan}
\end{equation}
where $\xi=2(\sigma^{\rm tot}_{pp}n_{p0}+\sigma^{\rm tot}_{pn}n_{n0}) a$
is the effective optical depth.
Note that in this case, the OLA of the Glauber model is strictly applicable 
when the effective optical depth is close to zero and the proton wave length 
is sufficiently short.  In the limit of complete absorption 
($\xi\to\infty$), however, Eq.\ (\ref{reacan}) reproduces the correct answer 
$\pi a^2$.  One may thus expect that the OLA provides a fairly good prediction 
of the reaction cross section even for intermediate values of $\xi$ as long as 
the incident energy is sufficiently high.  Nevertheless, 
Eq.\ (\ref{reacan}), if the exponential term in the right side works,
is at odds with the power-law $\sigma^{\rm tot}_{pN}$ dependence of the 
reaction cross section that is empirically suggested \cite{IL}.
Even if the exponential term is negligible, the energy dependence of the 
reaction cross section as deduced from Eq.\ (\ref{reacan}) shows only a
weak dependence on $A$, which is again at odds with the empirical behavior 
\cite{Auce}.

     We now consider the case in which $a_n$ and $a_p$ are generally
different from each other.  In this case, as we shall see, the reaction 
cross section (\ref{reaccs}) in the zero range limit
includes a term depending exponentially on 
the neutron skin thickness $a_n-a_p$.  In fact, when 
one retains corrections to the black-disk limit 
($\sigma_{pN}^{\rm tot}\to\infty$) by transparency of the skin 
region, the reaction cross section (\ref{reaccs}) in the zero range limit 
has a form
\begin{eqnarray}
  \sigma_{\rm R} &\simeq& \pi a_n^2 
\nonumber \\ & &
+2\pi\left[\frac{e^{-\zeta_n \sqrt{a_n^2-a_p^2}}}{\zeta_n^2}
\left(1+\zeta_n \sqrt{a_n^2-a_p^2}\right)-\frac{1}{\zeta_n^2}\right],
\nonumber \\ & &
\label{reacapprox}
\end{eqnarray}
where $\zeta_n=2\sigma^{\rm tot}_{pn}n_{n0}$.

     It is instructive to note that Eq.\ (\ref{reacapprox}) reduces to
$\pi a_n^2$ in the black-disk limit.  For large but finite values of 
$\zeta_n$, the remaining term in the right side of Eq.\ (\ref{reacapprox})
starts to play a role ahead of the corrections responsible for 
transparency of the inner region in which protons are present.  Interestingly, 
this term is a decreasing function of $a_n$ and in some cases acts to cancel 
an increase of the main term $\pi a_n^2$ by an increment of $a_n$.  As we
shall see, this term is responsible for the $L$ dependence of 
$\sigma_{\rm R}$ as shown in Figs.\ 1--3.

     Let us now assume that the distribution (\ref{ni}) roughly 
corresponds to a rectangular one discussed above.  Then, one can 
set the $L$ and $K_0$ dependence of $a_n$ and $a_p$ in such a way 
that $a_n$ increases linearly with $L$, while any other dependence
is negligibly weak (see Fig.\ 7 in Ref.\ \cite{OI}).  For $^{80}$Ni, 
in particular, we set $a_n$ and $a_p$ as 
\begin{equation}
a_n\approx 5.4+0.3 \left(\frac{L}{170~\rm{MeV}}\right)~\rm{fm},
\label{an}
\end{equation}
\begin{equation}
a_p\approx 5.1~\rm{fm},
\label{ap}
\end{equation}
in reference to the root-mean-square neutron and proton radii
calculated in Ref.\ \cite{OI}.  Here we have taken into account
a factor $\sqrt{5/3}$, which is the ratio of the half-density and
root-mean-square radii for rectangular distributions.
By using such a rough correspondence, we can understand the peculiar 
$L$ dependence of $\sigma_{\rm R}$ obtained in the OLA calculations.

     In fact, as shown in Fig.\ 4, we find that Eq.\ (\ref{reacapprox})
for various sets of $\sigma^{\rm tot}_{pn}$ gives a roughly
linear dependence on $L$, while this dependence is suppressed for
small values of $\sigma^{\rm tot}_{pn}$.  This is consistent with the tendency
seen in Figs.\ 1-3, although higher order corrections to the 
black-disk limit have to be significant for small values of $\zeta_n$ or,
equivalently, small $\sigma^{\rm tot}_{pn}$.

     Figures 1-4 suggest that the magnitude of the reaction cross sections
depends on the incident energy more remarkably in the OLA calculations
than in Eq.\ (\ref{reacapprox}).  This is because in the OLA calculations
we used the density distributions of form (\ref{ni}), which has a
surface diffuseness and thus allow reactions to occur in an even farther 
region than the radius of the rectangular distributions.

\begin{figure}[t]
\epsfig{file=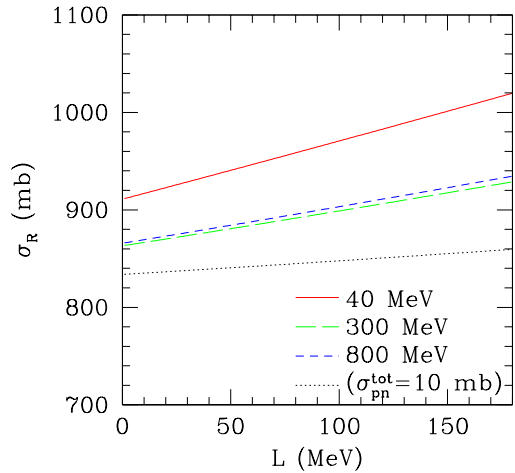,scale=0.4}
\caption{(Color online) The total reaction cross sections calculated from 
Eq.\ (\ref{reacapprox}) for $^{80}$Ni by using Eqs.\ (\ref{an}) and 
(\ref{ap}) as $a_n$ and $a_p$.  The full, long-dashed, and 
short-dashed lines are the results obtained for $\sigma_{pn}^{\rm tot}=220$, 
35, 38 mb, which are relevant at incident energies of 40, 300, 800 MeV 
\cite{PDG}. For comparison, we also plot by a dotted line the result obtained 
for $\sigma_{pn}^{\rm tot}=10$ mb.}
\label{rectang}
\end{figure}

     In summary, we have discovered that the OLA of the Glauber theory 
exhibits a nonnegligible exponential dependence of the total proton-nucleus 
reaction cross sections on the neutron skin thickness at incident energies
where $\sigma_{pn}^{\rm tot}$ is sufficiently small.  It is, however, 
important to note that the validity of using the OLA of the Glauber theory 
in evaluating the total reaction cross sections in the energy range
considered here is far from obvious.  We also note that the Coulomb force
ignored here has to play a role in distorting the proton trajectory at the 
lowest energy.  For duly describing the energy and size dependence of the 
total reaction cross sections, therefore, alternative approaches based on 
empirical data for the total reaction cross sections such as those in 
Refs.\ \cite{IL,IKO} might be useful.

\section*{Acknowledgments}

This work was supported in part by the Japan International Cultural 
Exchange Foundation (JICEF).


\begin{thebibliography}{99}
\bibitem{IOA} K. Iida, K. Oyamatsu, and B. Abu-Ibrahim, 
Phys. Lett. B {\bf 576}, 273 (2003).
\bibitem{OI} K. Oyamatsu and K. Iida, Prog.\ Theor.\ Phys.\ 
{\bf 109}, 631 (2003).
\bibitem{YCL} D.H. Youngblood, H.L. Clark, and Y.-W. Lui, 
Phys.\ Rev.\ Lett.\ {\bf 82}, 691 (1999).
\bibitem{Nat} J.B. Natowitz, K. Hagel, Y. Ma, M. Murray, L.Qin, R. Wada,
and J. Wang, Phys.\ Rev.\ Lett.\ {\bf 89},  212701 (2002).
\bibitem{Prakash} M. Prakash, T.L. Ainsworth, and J.M. Lattimer, 
Phys.\ Rev.\ Lett.\ {\bf 61}, 2518 (1988).
\bibitem{OI2} K. Oyamatsu and K. Iida, Phys.\ Rev.\ C
{\bf 75}, 015801 (2007).
\bibitem{O} K. Oyamatsu, Nucl.\ Phys.\ {\bf A561}, 431 (1993).
\bibitem{Glauber} R.J. Glauber, in {\it Lectures in Theoretical 
         Physics} (Interscience, New York, 1959), Vol. 1, p.\ 315.
\bibitem{bhks} B. Abu-Ibrahim, W. Horiuchi, A. Kohama, and 
Y. Suzuki, Phys. Rev. C {\bf 77}, 034607 (2008).
\bibitem{BS2}  A. Kohama, K. Iida, and K. Oyamatsu,
         Phys.\ Rev.\ C {\bf 72}, 024602 (2005).
\bibitem{IL} A. Ingemarsson and M. Lantz, Phys.\ Rev.\ C {\bf 72}, 
064615 (2005). 
\bibitem{Auce} A. Auce {\it et al.}, Phys.\ Rev.\ C {\bf 71}, 
064606 (2005). 
\bibitem{PDG}  From Particle Data Group via 
http://pdg.lbl.gov/current/xsect/.
\bibitem{carlson} R.F. Carlson, 
         At.\ Data Nucl.\ Data Tables {\bf 63}, 93 (1996).
\bibitem{IKO} K. Iida, A. Kohama, and K. Oyamatsu, 
J. Phys.\ Soc.\ Jpn.\ {\bf 76}, 044201 (2007).

\end{thebibliography}
\end{document}